\documentclass{article}
\usepackage{spconf,amsmath,graphicx}
\usepackage{xcolor}
\usepackage{booktabs}
\usepackage{hyperref}
\usepackage{caption}
\usepackage{tikz}
\usepackage{pifont}
\newcommand{\cmark}{\ding{51}}%
\newcommand{\xmark}{\ding{55}}%
\newcommand{\trillsson}{{TRILLsson}}
\newcommand{\wavvec}{{Wav2Vec$2.0$}}
\title{TRILLsson: Distilled Universal Paralinguistic Speech Representations}

\name{Joel Shor$^1$, Subhashini Venugopalan$^2$}
\address{Verily Life Sciences$^1$, Google Research$^2$ \\
joelshor@verily.com}

\begin{document}
%
\maketitle
\begin{abstract}
Recent advances in self-supervision have dramatically improved the quality of speech representations.
However, deployment of state-of-the-art embedding models on devices has been restricted due to their limited public availability and large resource footprint.
Our work addresses these issues by \textbf{publicly releasing a collection of paralinguistic speech models}\footnote[1]{\scriptsize\url{https://tfhub.dev/s?q=trillsson}} that are small and near state-of-the-art performance. Our approach is based on knowledge distillation, and our models are distilled on public data only. We explore different architectures 
and thoroughly evaluate our models on the Non-Semantic Speech (NOSS) benchmark.
Our largest distilled model is \textbf{less than 15\% the size} of the original model (314MB vs 2.2GB), achieves \textbf{over 96\% the accuracy on 6 of 7 tasks}, and is trained on 6.5\% the data. The smallest model is \textbf{1\% in size} (22MB) and achieves over \textbf{90\% the accuracy on 6 of 7 tasks}. Our models outperform the open source Wav2Vec 2.0 model on 6 of 7 tasks, and our smallest model outperforms the open source Wav2Vec 2.0 on both emotion recognition tasks despite being 7\% the size.
\end{abstract}
\section{Introduction}
\label{sec:intro}

Self-supervised learning for audio has improved the quality of speech representations, resulting in huge performance gains on downstream tasks~\cite{bigssl, hubert, baevski2020wav2vec}. 
However, a few obstacles prevent these representations from being widely adopted on devices, particularly for paralinguistic speech tasks which focus on aspects of speech other than the textual meaning.
First, recent self-supervised models are often extremely large, making them challenging to use in resource-constrained environments like mobile phones (e.g. HuggingFace's Wav2Vec 2.0~\cite{baevski2020wav2vec} at 360 MB).
Second, due to the private nature of most speech data, high-performance models are often not publicly released (e.g. CAP12~\cite{cap12}). In this work, we overcome both constraints using public data and knowledge distillation~\cite{distillation}. Specifically, we knowledge distill a recent state-of-the-art paralinguistic speech representation ``Conformer Applied to Paralinguistics" (CAP12~\cite{cap12}) into a series of models, which we call \trillsson.

Our approach primarily relies on ``knowledge distillation"~\cite{distillation}. We train small student models on several fixed-length input architectures, including ResNets, EfficientNets, and Transformers, to match the arbitrary-length input CAP12 (teacher) Transformer embeddings. Our architectures explore the model size versus performance tradeoff. To our knowledge, this is the first successful cross-architecture distillation from a Transformer to non-Transformer model using a different dataset. 

We use nearly 58K hours of publicly available speech data from Libri-light~\cite{librilight} and Audioset~\cite{hershey2017cnn} for distillation. 
For evaluation, we use the ``Non-Semantic Speech Benchmark" (NOSS)~\cite{trill}. NOSS includes 7 tasks such as emotion recognition and speaker identification that require slow-time features.
Additionally, we demonstrate the superior performance of \trillsson{} models by comparing them with existing publicly available representations such as TRILL~\cite{trill} and \wavvec~\cite{baevski2020wav2vec}.
Our contributions are:
\begin{enumerate}\itemsep0em
   \item Create generally-useful paralinguistic models that are small enough to run on-device.
   \item Demonstrate successful cross-architecture knowledge distillation from Transformers to fixed-context convolutional networks.
    \item Publicly release models at different points of the model size and performance trade-off curve.
    \item Identify the best paralinguistic representation in the public Wav2Vec2.0 model and demonstrate that our models outperform it.
 \end{enumerate}

\section{Background and related works}
\label{sec:related}
Self-supervised representation learning has shown remarkable success in vision~\cite{simclr} and speech
recognition~\cite{baevski2020wav2vec}. The \wavvec~\cite{baevski2020wav2vec} and Conformer models~\cite{conformer} are most relevant to our work. 
\cite{baevski2020wav2vec} was one of the first frameworks to successfully combine Transformers~\cite{transformer} and a self-supervised contrastive learning objective for speech. The same training objective was subsequently combined with Conformer architectures~\cite{conformer}, which added convolution filters to Transformer layers, producing further improvements in semi-supervised speech recognition applications~\cite{bigssl}. Recently,  \cite{cap12} developed Conformer-based models (CAP12) that created representations for
non-ASR speech analysis and paralinguistics tasks. Linear models on time-averaged CAP12 representations performed at or above state-of-the-at across several tasks simultaneously. However, as with most self-supervised models, their resource footprint (memory and compute) makes them less suitable for on-device applications. In this work, we distill the CAP12 model from \cite{cap12} to several ``lite'' architectures for use on mobile devices, and we release them publicly.

Distillation~\cite{distillation} has been popular for transferring knowledge from large models to smaller ones. We distill the Conformer model to a variety of smaller, fixed-length input architectures that have been used in audio classification, such as ResNets~\cite{hershey2017cnn}, EfficientNets~\cite{efficientnetv2}, and ASTs~\cite{ast}. Other works have applied distillation to speech representations~\cite{frill}, but to our knowledge this work is the first successful speech embedding distillation from arbitrary-length input Transformers to fixed-length input models.
\section{Approach}
\label{sec:experiments}


\begin{table}[!t]
\captionsetup{size=footnotesize}
\centering
\caption{Training datasets. We only use AudioSet and Libri-light for distillation. YT-U was used to train CAP12~\cite{cap12}. }\label{tab:train_ds}
\footnotesize
\begin{tabular}{c c c c c c } \toprule
  Dataset & Total (Hours) & Samples &  Avg len (s)  \\
  \midrule
 Speech AudioSet & 4.9K & 1.8M & 9.9 \\
 Libri-light & 53K & 6.0M & 31.8 \\
 \midrule
 YT-U & 900K & 295M & 11.0 \\
 \bottomrule
\end{tabular}
\end{table}

\begin{table}[!t]
\captionsetup{size=footnotesize}
\centering
\caption{Downstream evaluation datasets.
$^*$Results in our study used a subset of Voxceleb filtered according to YouTube's privacy guidelines.}\label{tab:eval}
\footnotesize
\begin{tabular}{c c c c c c } \toprule
  Dataset        & Target & Classes     & Samples     &  \begin{tabular}{@{}c@{}}Avg \\ length (s)\end{tabular}  \\
  \midrule
 VoxCeleb$^*$~\cite{voxceleb}  
 &  Speaker ID   & 1,251    & 12,052     & 8.4 \\
 VoxForge~\cite{voxforge}   
 & Language ID   & 6        & 176,438   & 5.8 \\
 \begin{tabular}{@{}c@{}}Speech \\ Commands\cite{speechcommands}\end{tabular} 
 & Command       & 12       & 100,503   & 1.0  \\
 ASVSpoof~\cite{asvspoof}   
 & \begin{tabular}{@{}c@{}}Synthetic \\ or not\end{tabular}
 & 2  & 121,461  & 3.2  \\
 Euphonia~\cite{euphonia}
 & Dysarthria & 5 & 15,224 & 6.4 \\
 CREMA-D~\cite{cremad}
 & Emotion       & 6        & 7,438     & 2.5 \\
 IEMOCAP~\cite{iemocap} 
 & Emotion       & 4        & 5,531         & 4.5    \\
 \bottomrule
\end{tabular}
\end{table}

\begin{figure}[t]
{\centering
  \includegraphics[width=0.9\columnwidth]{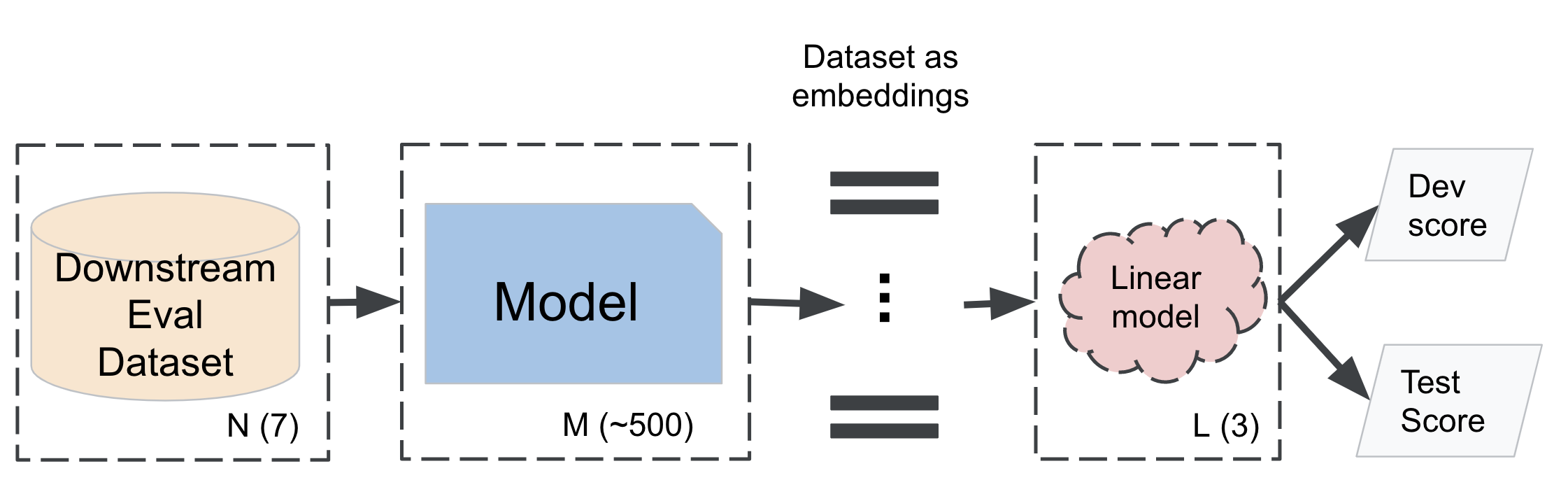}\hspace{0.2cm}
  \caption{Depiction of the evaluation process.}\label{fig:eval_flow}}
  \vspace{-3mm}
\end{figure}



Shor et. al. \cite{cap12} demonstrated that the 1024 dimension representation of the 12th layer of the CAP Conformer model (referred to as ``CAP12'') achieved at or near state-of-the-art performance across all tasks in the paralinguistic Non-Semantic Speech Benchmark (NOSS)~\cite{trill}. CAP12 is a 606M parameter (2.2GB) Conformer model trained via a modified \wavvec\  self-supervised training loss on a 900M+ hour speech dataset derived from YouTube (YT-U~\cite{bigssl}, Tab.~\ref{tab:train_ds}). 
In this work, we use CAP12 as the teacher and distill this model to several ``lite'' architectures based on the teacher-student distillation approach~\cite{distillation}.

\subsection{Student architectures.}
\label{sec:kd}
We explore 3 different student architectures of varying sizes.
\begin{enumerate}\itemsep0em
    \item \textbf{Audio Spectrogram Transformer (AST)}~\cite{ast} is a Transformer-based model for audio classification. We train student models with different depths and widths.
    
    \item \textbf{EfficientNetv2}~\cite{efficientnetv2} was designed by neural architecture search on image classification. The architecture is mobile friendly. Different versions of this architecture vary in terms of depths and filters.
    
    \item \textbf{Resnetish}~\cite{hershey2017cnn} are modified ResNet-50 architectures designed to take audio spectral features as input. Different versions of this architecture include different depths and different number of filters per layer.
\end{enumerate}

\begin{centering}
\begin{table*}[t]
\scriptsize
\vspace{-0.2cm}
\caption{Test performance on the NOSS Benchmark and extended tasks. 
``Prev. SoTA" are usually domain specific, but all other rows are linear models on time-averaged input. TRILLsson model sizes are shown without frontends.
$\uparrow$ indicates higher values are better, and $\downarrow$ indicates lower is better.
$^\dagger$We use a filtered subset of Voxceleb1 according to YouTube’s privacy guidelines. We omit previous SoTA results on this dataset, since they used the entire dataset.
$^{**}$ASVSpoof uses equal error rate~\cite{asvspoof}. We report the best single-model performance (as compared to model ensembles).
$^{\#}$Euphonia is the only non-public dataset. We use a larger dataset than was reported on in~\cite{cap12}.
$^*$We use the public Wav2Vec 2.0 model from \href{https://huggingface.co/docs/transformers/model_doc/wav2vec2}{Hugging Face}~\cite{transformers_lib}
}\label{tab:results}

\centering
\begin{tabular}{@{} llcc|ccccccc @{}}
\toprule[2pt]
\begin{tabular}{@{}c@{}}Model\\(input size)\end{tabular} &
\begin{tabular}{@{}c@{}}Params\\ (M)\end{tabular} &
\begin{tabular}{@{}c@{}}Size\\ (MB)\end{tabular} &
Public &
Voxceleb1$^\dagger$ $\uparrow$ & 
Voxforge $\uparrow$\ & 
\begin{tabular}{@{}c@{}}Speech $\uparrow$\\ Commands\end{tabular}   &  
\begin{tabular}{@{}c@{}}ASVSpoof\\ 2019$^{**}$ $\downarrow$ \end{tabular} & 
Euphonia$^{\#}$ $\uparrow$ &
CREMA-D $\uparrow$ &
IEMOCAP $\uparrow$
\\
\toprule[2pt]
\textbf{Prev. SoTA}
& -
& -
& -
& -
& 99.8~\cite{cap12} 
& 97.9~\cite{speech_commandssota} 
& 2.5~\cite{cap12} 
& -
& 88.2~\cite{cap12} 
& 79.2~\cite{cap12}
\\
\midrule
\textbf{CAP12} \\
\quad (full)  
& 606 & 2,200 & \xmark
& 51.0  & 99.7      & 97.1 & 2.5 & 46.9  & 88.2 & 75.5  \\
\quad (3s) 
& 606 & 2,200 & \xmark
& 47.9  & 99.4      & 97.1 & 3.8 & 46.9  & 88.1 & 74.3    \\
\quad (2s) 
& 606 & 2,200 & \xmark
& 48.1  & 99.4      & 97.0 & 6.9 & 46.9  & 85.3 & 72.7   \\
\midrule
\midrule
\textbf{Baselines} \\
\quad Wav2Vec2 Sm. L6$^*$
& 93.4 & 360 & \cmark
& 17.9 & 98.5 & \textbf{95.0} & 6.7 & 48.2 & 77.4 & 65.8 \\
\quad Wav2Vec2 Sm.$^*$
& 93.4 & 360 & \cmark
& 1.7 & 95.9 & 89.3 & 11.2 & 50.0 & 58.0 & 52.4 \\
\quad TRILL 
& 24.5 & 87 & \cmark
& 13.8 & 84.5 & 77.6 & 6.3 & 47.0 & 65.7 & 55.4 \\
\quad YAMNet
& 3.7 & 17 & \cmark
& 9.6 & 79.8 & 78.5 & 6.7 & 43.8 & 66.4 & 57.5 \\
\midrule
\textbf{TRILLsson} \\


\quad 5 (AST) 
& 88.6 & 314 & \cmark
& \textbf{46.2} & \textbf{99.7} & 93.9 & \textbf{5.4} & 48.1 & 86.1 & 72.7  \\


\quad 4 (AST) 
& 63.4 & 224 & \cmark
& 43.1 & 99.6 & 94.5 & 7.1 & \textbf{50.7} & \textbf{86.2} & \textbf{73.2}    \\


\quad 3 (EffNetv2) 
& 21.5 & 99 & \cmark
& 40.1 & 99.2 & 93.2 & 6.8 & 47.4 & 83.2 & 70.3    \\

\quad 2 (EffNetv2) 
& 8.1 & 42 & \cmark
& 37.5 & 99.2 & 92.1 & 6.6 & 44.6 & 82.6 & 69.8    \\

\quad 1 (ResNet) 
& 5.0 & 22 & \cmark
& 36.6 & 98.6 & 91.2 & 7.5 & 43.3 & 81.3 & 68.5    \\
\bottomrule
\end{tabular}
\vspace{-4mm}
\end{table*}
\end{centering}



Based on the Figure 1 lower in \cite{cap12} and the fact that some benchmark datasets have audio that's mostly less than 3 seconds (Tab.~\ref{tab:eval})  we focus on student architectures that process 2 seconds of audio at a time. When operating on audio less than 2 seconds, we symmetrically pad the audio around the end.

For the frontend, we use window length of 25ms and a hop length of 10ms. We use log-magnitude mel-frequency spectrograms with 80 bins ranging from 125 Hz to 7500 Hz. The frame width for all models is 2 seconds of audio, and we treat the frame advance between successive model patches as a hyperparameter. Since the model sees audio of exactly 2 seconds during training (see Sec~\ref{subsec:matching}), the frame advance hyperparameter can be explored after the model is fully trained.

\subsection{Distillation targets: global vs local matching}
\label{subsec:matching}
There are two paradigms for generating targets from a teacher that borrow ideas from distilling large vision models~\cite{caron2021emerging}. In both cases the audio is first chunked into context windows which the student processes. Then the student is trained to match a target embedding generated using the teacher model. In ``global matching", the target is the average teacher embedding from the entire un-chunked audio clip. In the ``local matching" paradigm, the target is generated by averaging the teacher's output on the same context window that the student sees.  In this work, we focus on ``local matching."

\subsection{Training datasets}
We perform distillation on two open source datasets which together contain about 58K hours of speech data (Tab. \ref{tab:train_ds}). \textbf{Audioset}~\cite{audioset} clips are collection from YouTube, so it represents a variety of settings and acoustic environments. We use the speech subset of this data, which yields approximately 5K hours. \textbf{Libri-light}~\cite{librilight} contains 60k hours of audio derived from open-source audio books in the LibriVox project. It is the largest publicly available, unlabeled semi-supervised dataset to date. We note that the CAP12 teacher model~\cite{cap12} is trained on \textbf{YT-U}~\cite{bigssl}, which is a 900K hour dataset derived from YouTube.


\begin{figure*}[t]
{\centering
  \includegraphics[width=2.0\columnwidth]{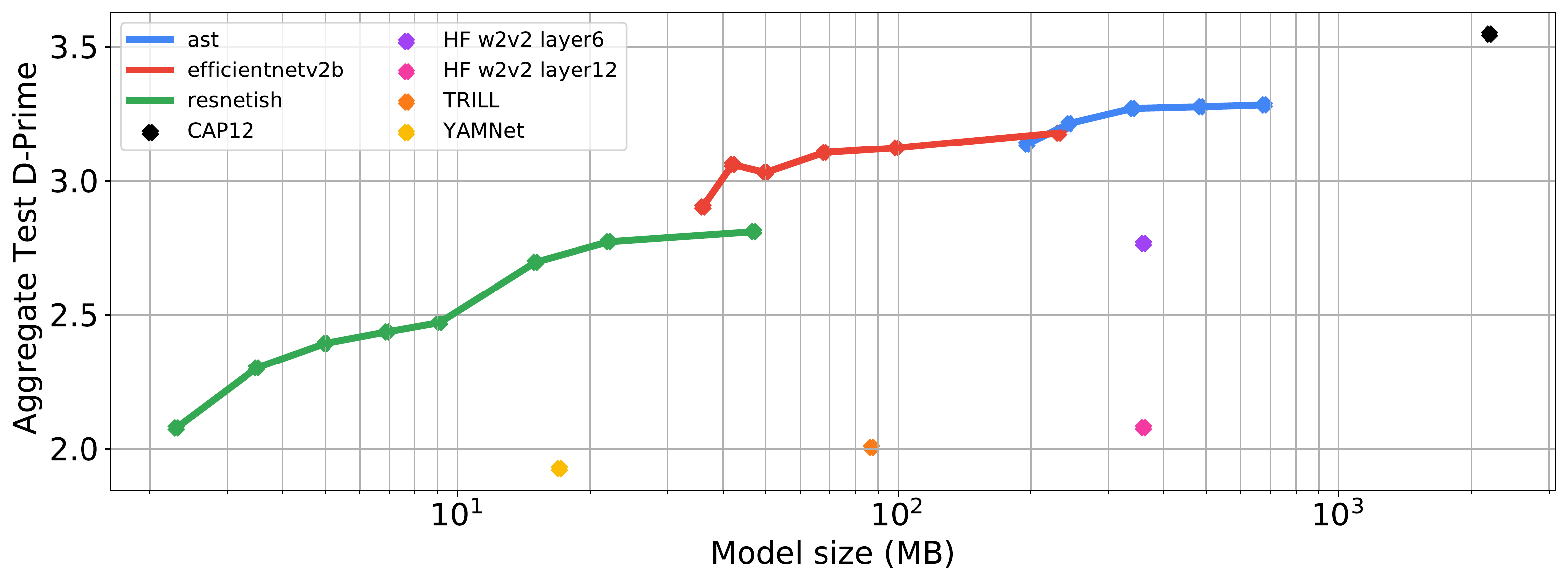}\hspace{0.2cm}
  \caption{Comparison of ``average $d'$" vs. ``model size" for various student model architectures and sizes. Performance in this figure is across test sets, although only dev-set performance is used to select "best" models.}\label{fig:performance}}
\end{figure*}
\vspace{-3mm}
\subsection{Representation Evaluation}
\label{subsec:eval}

We compare and evaluate the distilled representations on several tasks including detection of speaker, language, command, synthetic speech, dysarthria, and emotion. Table~\ref{tab:eval} provides an overview of the 7 datasets these tasks come from ~\cite{trill, cap12, asremb}.
As depicted in Fig~\ref{fig:eval_flow}, for each (model, eval dataset) pair, we first generate the candidate embeddings for the train, dev, and test splits. We then train three types of linear models on the train set embeddings. We take the model that performs best on the dev set, and report the model's performance on the test set as the score for that (model, eval dataset) pair. 

\vspace{-2mm}
\subsubsection{Aggregation metric: Equivalent D-Prime ($d'$)}
\label{subsubsec:dprime}
To fairly compare with previous results, we report on the typical performance metric for each dataset (accuracy or equal error rate). However, to aggregate performance into a single scalar, for the purpose of comparing embedding quality, we average the ``equivalent d-prime" ($d'$) metric, defined as follows: 
\begin{equation}
    d' = \sqrt{2}Z(AUC)
\end{equation}
where ``AUC" is the ``Area Under the Receiver Operating Characteristic Curve" (ROC AUC), and ``$Z()$" is the inverse Cumulative Distribution Function for the standard normal distribution.

$d'$ is better suited to aggregate performance on multiple datasets for two reasons. First, ROC AUC and $d'$, unlike accuracy, take into account performance at various levels along the accuracy/recall curve. This makes it more reflective of the overall performance of a particular representation. Second, unlike ROC AUC, $d'$ doesn't saturate in the highly performant regime. Thus, 1 unit of $d'$ is in some sense more equivalent, so averaging $d'$ values across datasets is more natural than averaging AUC values. We use the average $d'$ scores to sort and select models.

\subsection{Models for comparison}
\label{subsubsec:baselines}
To contextualize performance on the NOSS benchmark, we compare our models to 1) previous state-of-the-art (SoTA) results, 2) publicly available speech representation models, and 3) CAP12 with different input sizes. Previous SoTA models are been mostly domain-specific. The public models we compare to are:

\begin{enumerate}\itemsep0em
    \item \textbf{Hugging Face's Wav2Vec \text{2.0}}~\cite{baevski2020wav2vec,transformers_lib}: This model was trained on the approximately 1K hour Librispeech~\cite{librispeech}. We use the TensorFlow model from the Hugging Face library. We compute the performances for each layer, and find that \textbf{layer 6 of 11 performs the best overall}.
    \item \textbf{TRILL}~\cite{trill}: A Resnet triplet-loss model trained on AudioSet. We access this model from TensorFlow Hub.
    \item \textbf{YAMNet}~\cite{hershey2017cnn}: This is a supervised audio classification network, also trained on AudioSet. We use layer 19 as in \cite{trill}. We access this model through TensorFlow Hub.
\end{enumerate}

\section{Results}
\label{sec:results}

Table~\ref{tab:results} shows the sizes and performances of 5 \trillsson{} models as compared to CAP12 and other publicly available models. We see that the two largest \trillsson{} models outperform all publicly available models on all datasets except Wav2Vec2 layer 6 on speech commands, and that even \trillsson{}1, the smallest model, outperforms previously available embeddings on emotion recognition tasks. In particular, \trillsson{}2 outperforms the best Wav2Vec 2.0 representation on 5 of 6 public tasks despite being less than 12\% the size.

Compared to CAP12, the distilled models maintain performance on language identification and dysarthria detection. \trillsson{} models suffer minor degradation on speech emotion recognition tasks, and major degradation on speaker identification, speech commands, and fake speech detection.

Fig~\ref{fig:performance} shows the relationship between size and performance across different architectures. We see that Resnetish models perform best at the low end, EfficientNets perform best at sizes between 40-230MB, and AST performs best when larger then 230MB.

\textbf{Training on AudioSet improves performance on emotion recognition tasks:} Using a dependent t-test on paired samples between models trained on Libri-light and AudioSet versus just Libri-light, we observed statistically significant improvements in dev and test set accuracies on the speech emotion recognition tasks (CREMA-D and IEMOCAP, $n=86$, $p<0.03$). Interestingly, training on the combined dataset hurt performance on ASVSpoof2019 test set compared to either dataset on its own ($n=86$, $p<10^{-3}$), but the paired t-test did not reject the null hypothesis on the ASVSpoof2019 dev set.

\textbf{Best model curve robust to removing a single benchmark task:} Removing any single task keeps the size versus performance curve relatively stable. Either \trillsson{}5 or 4 are the best models with any single task removed. \trillsson{}3 and 2 appear in every optimal curve. \trillsson{}1 appears in every optimal curve unless 'voxforge' is removed.

\begin{centering}
\begin{table}[t]
\scriptsize
\vspace{-0.2cm}
\caption{Kendall rank coefficient between model orderings according to different criteria. For all computations, n = 435.}\label{tab:kendeltau}
\vspace{-2mm}
\centering
\begin{tabular}{@{} c|cccc @{}}
           & $d'$ Dev. & $d'$ Test      & Acc. Dev & Acc. Test \\
\midrule
$d'$ Dev  &   -         &     -        &    -     &     -    \\
$d'$ Test &   0.898     &     -        &  -       &    -      \\
Acc Dev &   0.893     &   0.854      &   -      &    -      \\
Acc Test&   0.860     &   0.821      &  0.820   &    -      \\
\midrule
\end{tabular}
\vspace{-5mm}
\end{table}
\end{centering}

\textbf{$d'$ orderings}: Table~\ref{tab:kendeltau} justifies our choice of $d'$ on the dev set as our ordering criteria. The Kendall rank coefficient, a correlation statistic for different orderings, shows that average $d'$ on the dev sets more closely correlates to average accuracy and average $d'$ than average accuracy on the dev set.




\vspace{-3mm}

\section{Conclusion}
\label{sec:conclusion}
In this work, we demonstrate that it's possible to distill huge models trained on large datasets to obtain much smaller models that perform well on paralinguistic speech tasks.
The distillation uses only \textbf{7\% of the training data} and is entirely from public sources. The models we obtain are between 22MB and 314MB, and achieve between \textbf{90\% and 96\% of the larger CAP12 accuracy on 6 of 7 tasks}. These models are between \textbf{1\% and 15\% the size} of the original model. We release the model to allow the research community to benefit from the practical applications of self-supervised representations for paralinguistic speech.

\vspace{-3mm}
\section{Acknowledgements}
We want to thank Aren Jansen, Channing Moore, Wei Han, Daniel Park, Yu Zhang, and the entire author list of \cite{bigssl}, without which this work wouldn't have been possible.

\bibliographystyle{IEEE}
\bibliography{biblio}


\end{document}